\documentclass[11pt]{article}
\usepackage{amssymb}
\usepackage{amsmath}
\usepackage{amsthm}
\usepackage{cancel}
\usepackage{slashed}
\usepackage{fullpage}
\usepackage{color}
\usepackage{braket}
\usepackage{graphicx}
\usepackage[small]{subfigure}
\usepackage{multirow}
\usepackage{verbatim}
\usepackage{cite}
\usepackage{bigints}
\usepackage{simplewick}
\usepackage{authblk}
\usepackage{hyperref}
\usepackage{empheq}
\usepackage{tikz-feynman}
\usepackage{tikz}

\newcommand{\bea}{\begin{eqnarray}}
\newcommand{\eea}{\end{eqnarray}}

\newcommand{\gsim}{\lower.7ex\hbox{$\;\stackrel{\textstyle>}{\sim}\;$}}
\newcommand{\lsim}{\lower.7ex\hbox{$\;\stackrel{\textstyle<}{\sim}\;$}}

\hypersetup{
    linktocpage,
    colorlinks,
    citecolor=darkgreen,
   linkcolor= darkgreen,
   urlcolor=darkgreen
}

\definecolor{darkred}{rgb}{0.5,0.0,0.0}
\definecolor{darkblue}{rgb}{0.0,0.0,0.9}
\definecolor{darkerblue}{rgb}{0.0,0.0,0.5}
\definecolor{darkgreen}{rgb}{0.0,0.5,0.0}
\definecolor{black}{rgb}{0.0,0.0,0.0}
\definecolor{brown}{rgb}{0.6,0.4,0.2}
\numberwithin{equation}{section}

\title{Collider search of light dark matter model with dark sector decay}

\author[a,b]{Yu Cheng\thanks{chengyu@mail.ecust.edu.cn}}
\author[a]{Wei Liao\thanks{liaow@ecust.edu.cn}}
\author[c,d]{Qi-Shu Yan}

\affil[a]{\small School of Physics, East China University of Science and Technology, Shanghai 200237, China}
\affil[b]{\small Tsung-Dao Lee Institute, and School of Physics and Astronomy, Shanghai Jiao-Tong University, Shanghai, China}
\affil[c]{\small School of  Physical Sciences, University of Chinese Academy of Sciences, Beijing 100049, China}
\affil[d]{\small Center for Future High Energy Physics, Institute of High Energy Physics, Chinese Academy of Sciences, Beijing 100049, China}

\begin{document} 
\maketitle

\begin{abstract}
 We explore the possibility that the dark matter relic density is not produced by thermal mechanism directly, 
but by the decay of other heavier dark sector particles which on the other hand can be produced by the thermal freeze-out mechanism.
Using a concrete model with a light dark matter from dark sector decay, we study the collider signature of the dark sector particles
in association with Higgs production processes. We find that the future lepton colliders can be a better place to probe the signature of this kind of light dark matter model than the hadron collider such as LHC. Meanwhile, it is found that a Higgs factory with center of mass energy 250 GeV has a better potential to resolve the signature of  this kind of light dark matter model than the Higgs factory with center of mass energy 350 GeV.
\end{abstract}

\section{Introduction}

The present direct and indirect search of Dark Matter(DM) have so far given a lot of constraints on the DM interaction
\cite{Aprile:2018dbl,PandaX-4T:2021bab,cui2017dark,Wang:2020coa,Akerib:2016vxi,Campos:2017odj,Aartsen:2014hva,Khachatryan:2014rra,ATLAS:2012ky}.
In particular, the direct detection experiments~\cite{Aprile:2018dbl,PandaX-4T:2021bab,cui2017dark,Wang:2020coa} have pushed  
the DM-nucleus interaction cross section to be less than around $10^{-46}$ cm$^2$ for GeV-TeV scale DM particles.
These direct detection measurements have raised serious concerns that the GeV-TeV scale DM should have interactions much weaker 
than we thought, such that for the DM candidate of weakly interacting massive particle, 
an over-production of the relic density in thermal production mechanism seems hard to avoid.

One way out of this difficulty is to note that the dark sector particle directly involved in the thermal production mechanism
may not be the DM candidate, but rather the mother particle of a decay process with the DM particle as the children particle.
In such a scenario, the DM relic density is derived from the relic density of another dark sector particle obtained from the thermal production mechanism. 
Since the mass of the DM particle could be much smaller than the mass of other dark sector particle,
 one can obtain a relic density of DM much smaller than the relic density directly obtained in the thermal production mechanism. 
Therefore, in such a scenario, one can still have a weak scale interaction between the dark sector and the visible sector, 
and meanwhile obtain a DM relic density consistent with cosmological observations.

This possibility of light DM from dark sector decay has been explored in a previous work~\cite{CL2020} . In this article, we further study some collider phenomenology of this scenario of DM. 
We will focus on the potential collider signature of a DM model which can have light DM from dark sector decay.
We study the feasible signatures at future lepton colliders, such as CEPC, FCC-ee and ILC. For the sake of comparison, we also study the signature and constraints from the hadron collider such as the LHC.

The article is organized as follows. 
In the next section, we quickly review the DM model we are considering and the various constraints on the model.
Then we study the signature of the DM model at future lepton colliders.
After a detailed analysis of signature on the lepton collider, we give a brief analysis on the signature on the hadron collider.
We summarize in the last section.
 
\section{The model of light DM and dark particle search at colliders}
\label{sec:model}
 The model we consider is an extension of  the Type-I seesaw model where the  right-handed neutrinos $N_{R}$ are coupled to a dark sector 
consisting of a real scalar $\phi$ and a Dirac fermion $\chi$~\cite{CL2020}. 
The model is similar to DM models  in which the $N_{R}$ works as a  portal to the DM\cite{Macias:2015cna, Blennow:2019fhy,Escudero:2016ksa,Gonzalez-Macias:2016vxy,Baek:2013qwa}. 
Previous works on this type of models have focused on the possibility that the DM relic density 
is achieved by the thermal freeze-out mechanism\cite{Escudero:2016ksa,Gonzalez-Macias:2016vxy,Bandyopadhyay:2018qcv} or by the freeze-in mechanism\cite{Chianese:2018dsz,Bandyopadhyay:2020qpn,Chianese:2019epo,Chianese:2020khl}.
Here, we consider the possibility that the DM relic density is given mainly by the decay of other heavier particles in the dark sector.

We impose a $Z_2$ symmetry in the Lagrangian with the dark sector particles $\chi$ and $\phi$ odd and SM fermions even under the $Z_2$ operation. 
This guarantees the lightest of the dark sector particles to be stable, and makes it a DM candidate. 
We can write the full Lagrangian as the sum of four parts
\begin{equation}
\mathcal{L}=\mathcal{L}_{\mathrm{SM}}+\mathcal{L}_{\text {Seesaw }}+\mathcal{L}_{\mathrm{DS}}+\mathcal{L}_{\text {int}}
\end{equation}
where the first term is the Standard Model(SM) Lagrangian and the other three terms are as follows.
\begin{eqnarray}
\mathcal{L}_{\text {Seesaw }}&&=-Y_{\alpha \beta} \overline{L^{\alpha}} \tilde{H} N_{R \beta}-\frac{1}{2} M_{N} \overline{N_{R}^{c}} N_{R}+\text { h.c. } \\
\mathcal{L}_{\text {DS }}&&=\bar{\chi}\left(i \not\partial-m_{\chi}\right) \chi + \frac{1}{2} \partial_{\mu} \phi \partial^{\mu} \phi-\frac{1}{2}\mu_{\phi}^{2} \phi^{2} -\frac{1}{4}\lambda_{\phi} {\phi}^4 
-\lambda_{H \phi}\left(H^{\dagger} H\right)\phi^{2} .\\
\mathcal{L}_{\text {int }} &&=-\left( y_{\text {DS }} \phi \bar{\chi} N_{R}+h.c \right).
\end{eqnarray}
In these equations, $L^\alpha$ is the left-handed lepton doublets, $H$  the SM Higgs doublet with $\tilde{H}=i \tau_{2} H^{*}$,
$m_{\chi}$ the mass of $\chi$, $M_N$ the mass of the right-handed neutrinos. 
There are three generations of right-handed neutrinos and we consider the simplest case where three generations of right-handed neutrinos are degenerate.
After the electroweak symmetry breaking, the scalar $\phi$ gets a further contribution to its mass and
we have 
\begin{equation}
m^2_{\phi} = {\mu}^2_{\phi}+\lambda_{H \phi} v_{ew}^2,
\end{equation} 
 where  $v_{ew}/\sqrt{2}$ is the vacuum expectation of Higgs doublet $H$ after the electroweak symmetry breaking.

As shown in \cite{CL2020}, the right relic abundance of DM can be obtained in this model by one scenario that the dark fermion $\chi$ serves as the DM candidate and is produced mainly through the decay of heavier dark scalar $\phi$. In this scenario, the coupling  $y_{DS}$ is assumed to be very small, typically at order $10^{-12}$. For such a small value of $y_{DS}$, the fermion $\chi$ is never in thermal equilibrium. We also assume that the coupling $\lambda_{H \phi}$ is not so small so that the scalar $\phi$ is in the thermal equilibrium with the SM particles at high temperature. As temperature drops down to the mass scale of $\phi$,
the heavier $\phi$ freezes out and decouples from the thermal bath. Then, $\phi$ after freeze-out decays into fermion $\chi$ and the right-handed neutrinos. The right-handed neutrinos would later decay to leptons and other particles, and we are left with a relic density of $\chi$ which is to be compared with the DM relic density found in cosmological observations. 

\begin{figure}[!t]
	\centering
	\subfigure[\label{fig:darkmattercons}]
	{\includegraphics[width=0.5\textwidth]{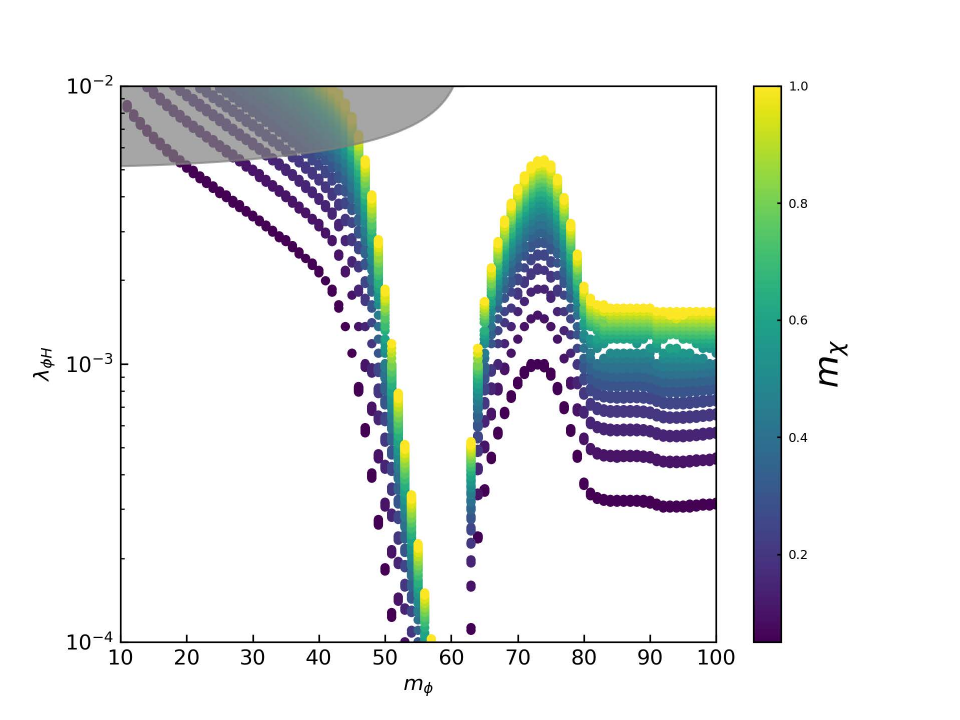}}
	\subfigure[\label{fig:darkphiphiALL}]
	{\includegraphics[width=.48\textwidth]{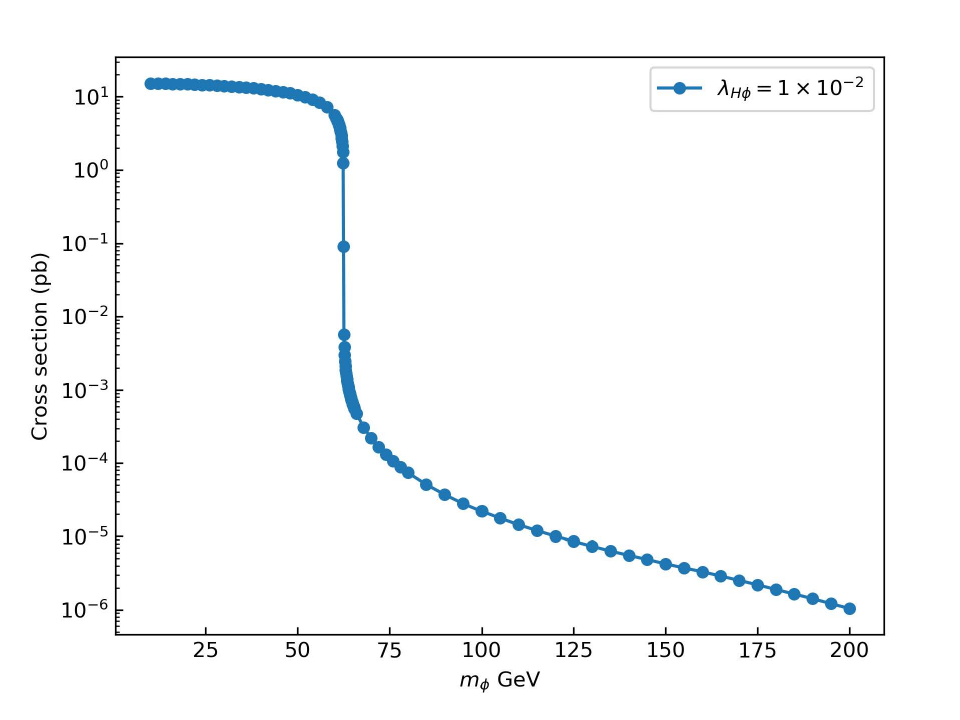}}   
	\caption{ (a) The allowed parameter space of $m_{\phi}$ and $\lambda_{H \phi}$ for $m_{\chi}$ in the range from 5 MeV to 1000 MeV. $M_{N} = 10$ GeV  and $y_{DS} = 10^{-12}$. The gray region is excluded by Higgs invisible decay.
		(b) Cross section for gluon gluon fusion production of $\phi$ pairs at pp collider, here we choose $\lambda = 0.01$ and 
		center-of-mass energy 13 TeV.
	} 
\end{figure}

In the previous work  \cite{CL2020}, only the case for $m_\phi > 200$ GeV was considered for obtaining the right relic density. 
The right relic density of DM can also be obtained for smaller $\phi$ mass. 
In this article, we consider here the parameter space with $m_{\phi} < 100$ GeV. 
The allowed parameter space of $m_{\phi}$ and $\lambda_{H \phi}$ for $m_{\phi}$ below 100 GeV is shown in Fig.~\ref{fig:darkmattercons}. 
We can see that we can obtain the right relic abundance,  and $\lambda_{H \phi}$ is demanded to be in the range $10^{-4} - 10^{-2}$ for $m_{\phi}$ 
in the range 10-100 GeV. 
Here we choose $y_{DS} = 10^{-12}$, $m_{N} = 10$ GeV, and $m_{\chi}$ in the range 5-1000 MeV. 
Note that for these parameters, the $\phi$ can decay sufficiently fast so that the DM relic density can be produced 
at sufficient high temperature, as can be shown similar to the analysis given in \cite{CL2020}. Typically for $m_{\phi} = 50$ GeV and $m_{\chi} = 1$ GeV with chosen parameters, the lifetime of $\phi$ is around 0.7s which would not give much impact on the BBN processes. 
We note that  the peak at around 75 GeV is due to the opening of the $\phi \phi \rightarrow W^{+} W^{-}$ annihilation channel 
and the numerical results shown in the figure is obtained by using the public code \textbf{MicrOmegas}\cite{Belanger:2013oya,Belanger:2018ccd} and \textbf{CALCHEP}\cite{Belyaev:2012qa} to solve the Boltzmann equation and calculate relic abundance. 
In numerical calculation, all the contributions from $2 \rightarrow 2$ process have been taken into account. 
The model implementation for \textbf{MicrOmegas} was done using \textbf{FeynRules}\cite{Alloul:2013bka} package.

For $m_{\phi} < m_{H}/2$, Higgs would decay to $\phi$ and would contribute to the Higgs invisible decay. 
The experimental data will give an additional constraint on model parameters $\lambda_{H \phi}$ and $m_{\phi}$. 
The branching ratio of Higgs invisible decay is demanded by latest experiment\cite{ATLAS:2020kdi} to be less than 11\%. 
In our model 
\begin{equation}
\Gamma_{h \rightarrow \phi \phi}=\frac{\lambda_{H \phi}^{2} v_{e w}^{2}}{8 \pi m_{H}} \sqrt{1-\frac{4 m_{\phi}^{2}}{m_{H}^{2}}}
\end{equation}
Only the parameters space for $\Gamma_{h \rightarrow \phi \phi} / (\Gamma_{S M}+\Gamma_{h \rightarrow \phi \phi}) \leq 11\%$ is allowed\cite{ATLAS:2020kdi}.
The gray region in Fig.~\ref{fig:darkmattercons} is excluded by the constraint from the branching ratio of Higgs invisible decay 
which is demanded to be less than 11\%. 

The dark particle $\chi$ does not couple to SM sector directly, so it's hard to find relevant signals of $\chi$ on the collider. But the scalar $\phi$ in the dark sector couples to the SM sector directly with a not so small coupling constant $\lambda_{H\phi}$, which gives us an opportunity to search this dark scalar $\phi$ at colliders. In general, the production of this dark sector particle $\phi$ will lead to a large $E_{\mathrm{T}}^{\mathrm{miss}}$ which can be a good signal for dark $\phi$ at colliders. Note here that the dark sector particle $\phi$ is a long-lived particle due to the very small coupling  $y_{DS}$, which can lead to a displaced vertex signal in the collider detector or long baseline experiment such as MATHUSLA\cite{MATHUSLA:2019qpy}, we will leave this topic in the future work.

\begin{figure}[!t]
	\setlength{\abovecaptionskip}{-20pt}
	\centering
	\subfigure[\label{fig:ggFphi}]
	{\includegraphics[width=.38\textwidth]{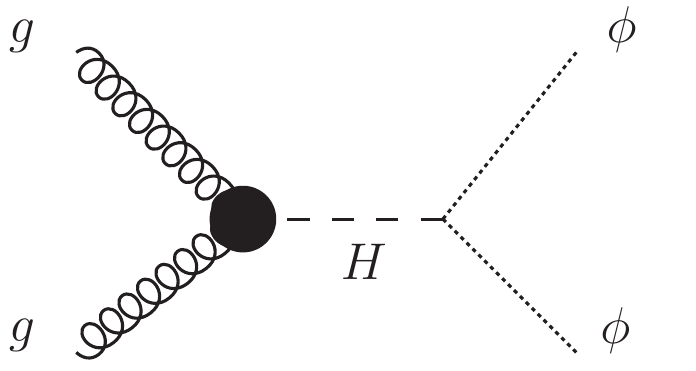}}
	\subfigure[\label{fig:EEZHphi}]
	{\includegraphics[width=.38\textwidth]{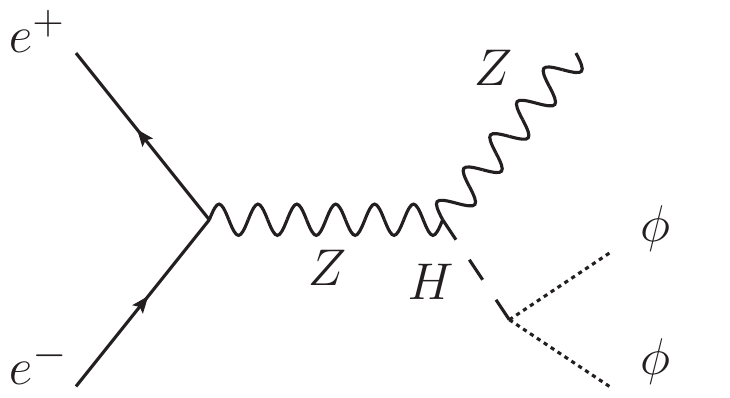}}
	\vspace{1cm}
	\label{Phiproduction}
	\caption{Major production processes of $\phi$ at pp and $e^{+} e^{-}$ colliders.}
\end{figure}

Dark sector particle $\phi$ is mainly produced by the gluon gluon fusion process through Higgs at hadron colliders, 
or by  the ZH($\rightarrow \phi \phi$) process at future Higgs factory which are shown in Fig.~\ref{Phiproduction}. The cross section for gluon gluon fusion production at hadron colliders can be obtained numerically by \textbf{MadGraph5\_aMC@NLO}\cite{Alwall:2011uj,Alwall:2014hca} package and the result is shown in Fig.~\ref{fig:darkphiphiALL} for 
$\lambda_{H \phi} = 0.01$ and center-of-mass energy $\sqrt{s} = 13$ TeV. The total cross section has been rescaled to 48.61 pb which is calculated at next-to-next-to-next-to-leading order (N3LO) in QCD\cite{Cepeda:2019klc}. We can see in Fig.~\ref{fig:darkphiphiALL} that the cross section drops sharply around $m_{\phi} = m_{H}/2 \simeq 62.5$ GeV.
 This is because for $m_\phi \lsim 62$ GeV, the production process is basically a $2 \to 1 $ process with the Higgs produced on-shell 
and then decays to $\phi \phi$, as can be seen in the left panel in Fig.  \ref{Phiproduction}, 
while for  $m_\phi \gsim 63$ GeV on-shell Higgs can not decay to $\phi \phi$
and the production of $\phi \phi$ shown in Fig.  \ref{Phiproduction} is mediated by off-shell Higgs.
So there is a transition from a $2 \rightarrow 1$ gluon gluon fusion production process to a $2 \rightarrow 2$ production process at around  $m_{\phi}\simeq 62.5$ GeV.
We can see in Fig.~\ref{fig:darkphiphiALL} that the cross section for $m_{\phi} > m_{H}/2$ is around 
five orders of magnitude smaller than that for $m_{\phi} <  m_{H}/2$. For the integrated luminosity at $300 fb^{-1}$, 
it is estimated that there are only few pairs of $\phi$ generated for $m_{\phi} > m_{H}/2$ at the LHC. 
For designed future $e^{+} e^{-}$ colliders, a rapid decrease of Higgs production cross section will also happen when $m_{\phi}\simeq 62.5$ GeV
which is associated with a transition from a $2 \rightarrow 2$ production process to a $2 \rightarrow 3$ production process.
Also for $e^{+} e^{-}$ collider running at 250-350 GeV, the energy conservation will suppress the production of the interested events with a $\phi$ mass much heavier than $m_{H}$/2. Therefore, in this work, we mainly focus on the parameter region with $m_{\phi} <  m_{H}/2$.

\section{Dark sector search via the Higgstrahlung process at $e^{+} e^{-}$ colliders}
\label{eecollider}
For the planned future Higgs factories like CEPC,FCC-ee and ILC with an integrated luminosity at the order of $ab^{-1} $, around $10^{6}$ Higgs bosons can be produced, which offers us an opportunity to search for the dark particle signal. 
In our model, the major signal process is $e^+ e^- \to Z H $ with the Higgs boson decaying into a pair of $\phi$ which can fly off the detectors as missing energy. Below we provide a feasibility study for $2 \ell + {E \!\!\!\!\slash}$ final states at $e^{+} e^{-}$ colliders with a detailed MC analysis.

Before presenting our MC analysis, one remarkable aspect of $e^{+} e^{-}$ colliders is worthy of highlighting, i.e. the full four momentum of the missing energy can be precisely reconstructed at $e^{+} e^{-}$ colliders. In contrast, only transverse momentum of the missing energy can be reconstructed at pp colliders. 

Meanwhile, another remarkable feature is that the charged leptons $\ell$ both in the final states and in the initial states can be measured accurately, which is crucial to reconstruct the mass parameters of mother particles in the signal and background events. In principle, our analysis can be extended to 2 jets plus missing energy case. Due to the large errors in the jet energy reconstruction, in this work we restrict to show the 2 charged lepton plus missing energy case. 

\begin{figure}[!htbp]
	\setlength{\abovecaptionskip}{-20pt}
	\centering
	\subfigure[\label{fig:SIGEEZH}]
	{\includegraphics[width=.40\textwidth,height=.18\textheight]{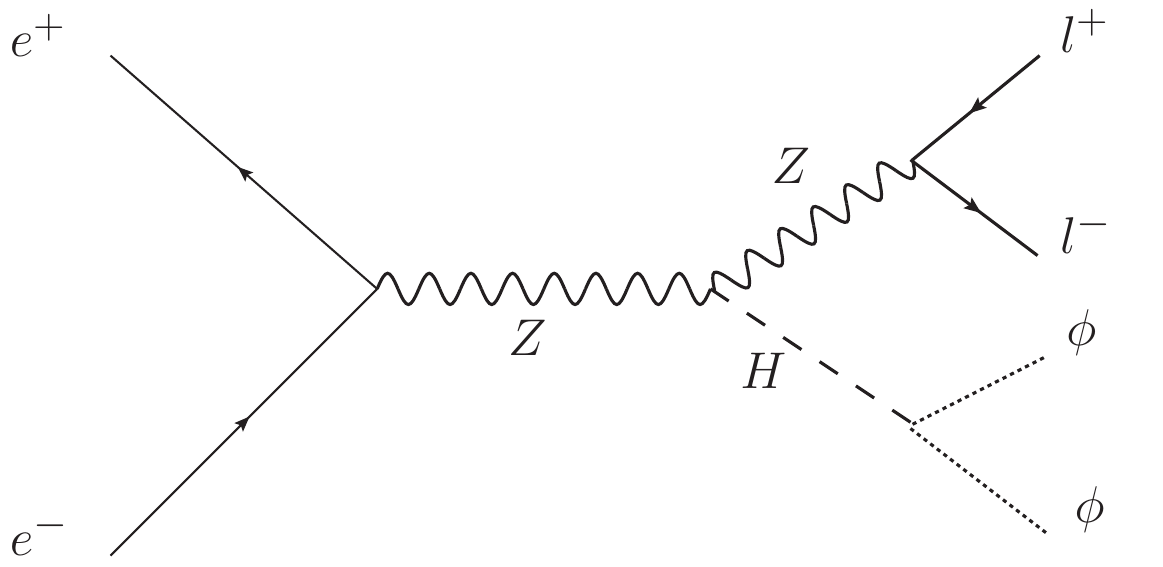}}
	\subfigure[\label{fig:BACEEZH}]
	{\includegraphics[width=.30\textwidth]{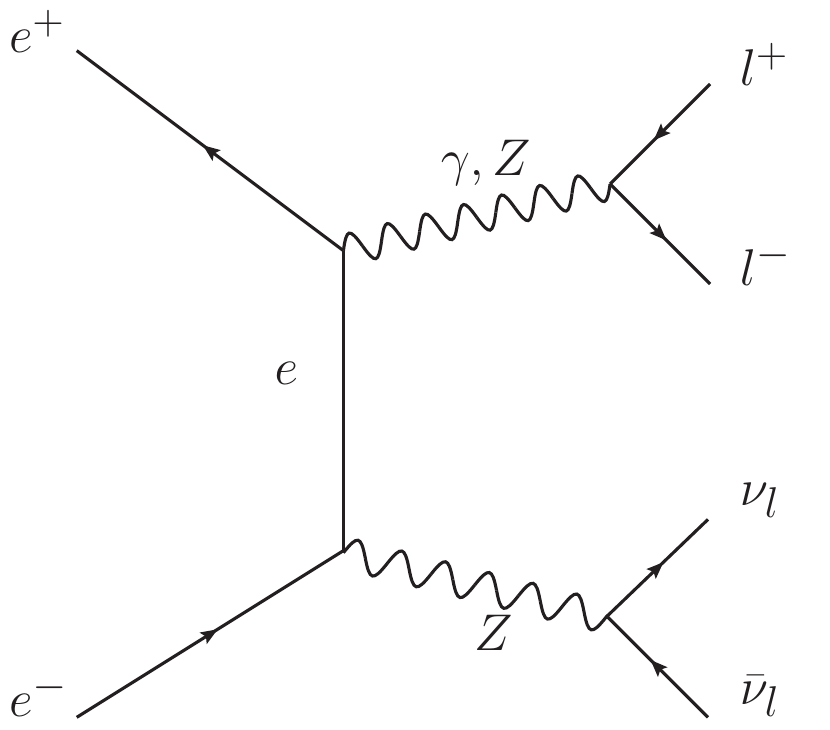}}
	\vspace{1cm}
	\label{EEZHdiagram}
	\caption{(a) Signal process at Higgs factory; (b)Major background process at Higgs factory.}
\end{figure}

In this work, both signal and background events at parton level are generated by using  the \textbf{MadGraph5\_aMC@NLO}\cite{Alwall:2011uj,Alwall:2014hca}. The main Feynman diagrams for signal and background processes are displayed in Fig.  \ref{EEZHdiagram}. The signal process and the main background process with decays are listed below 
 \begin{equation}
 \begin{aligned}
 &e^{+} e^{-} \rightarrow ZH \rightarrow l^{+} l^{-} + E^{\mathrm{miss}}(\phi \phi) \quad (\textbf{signal process})\,,\\
 &e^{+} e^{-} \rightarrow ZZ \rightarrow l^{+} l^{-}  + E^{\mathrm{miss}}(\nu \bar{\nu}) \ \quad (\textbf{background process})\,,
 \end{aligned}
 \end{equation}
where the missing energy in the signal process originates from Higgs boson's invisible decay, 
while the missing energy in the background process comes from Z boson decay.

To fully utilize the experimental precision and to generate the most relevant events, we introduce the following geometric acceptance cuts at the parton level. The leptons in the final state must be energetic, be near the central region of the detectors, and be isolated, i.e. $p_{T}(l) > 20$ GeV, $|\eta(\ell)|<2.5$, and $R(\ell)>0.4$. The leptons' momentum and energy in the non-central region has worse accuracy and it is better to neglect those events with leptons at the non-central region.

All acceptance cuts for the final state $\ell \ell+E^{\text {miss }}$ are provided in Table \ref{Cutees}. Such acceptance cuts are introduced so as to measure the full momentum of leptons in the final state precisely and to guarantee the successful reconstruction of the masses of mother particles of missing children particles. 

 \begin{table}[!htbp]
	\centering
	\setlength{\tabcolsep}{2.8mm}{
	\begin{tabular}{lllllllll}
	\multicolumn{9}{c}{Geometric Acceptance Cuts}    \\ \hline
	\multicolumn{3}{l}{Energetic Charged lepton} & \multicolumn{6}{l}{$P_{T}(l) > 20$ GeV } \\ \hline
	\multicolumn{3}{l}{Central Region} & \multicolumn{6}{l}{$|\eta(\ell)| < 2.5$ } \\ \hline
	\multicolumn{3}{l}{Isolation condition } & \multicolumn{6}{l}{$\Delta R_{\ell \ell} > 0.4$}\\  \hline 
%		\multicolumn{3}{l}{$m_{\ell \ell}$} & \multicolumn{6}{l}{$76<m_{\ell \ell}<106$ GeV} \\
	\end{tabular} 	
\caption{Geometric acceptance cuts for the $2 \ell+E^{\text {miss }}$ final states search at Higgs factory are shown. }	
	\label{Cutees}
}
\end{table}

For signal events, we choose two benchmark points with theoretical parameters (BP1, $m_{\phi} = 30$ GeV, $\lambda_{H \phi} = 0.005$) and (BP2, $m_{\phi} = 50$ GeV, $\lambda_{H \phi} = 8 \times 10^{-4}$) as two show cases to demonstrate how the recoil mass method can work to separate signal and background events. It should be mentioned that both of these two benchmark points can lead to the correct relic density for a cosmological DM candidate.

\begin{figure}[!htbp]
	\centering
	\subfigure[\label{fig:run5e330250}]
	{\includegraphics[width=.45\textwidth]{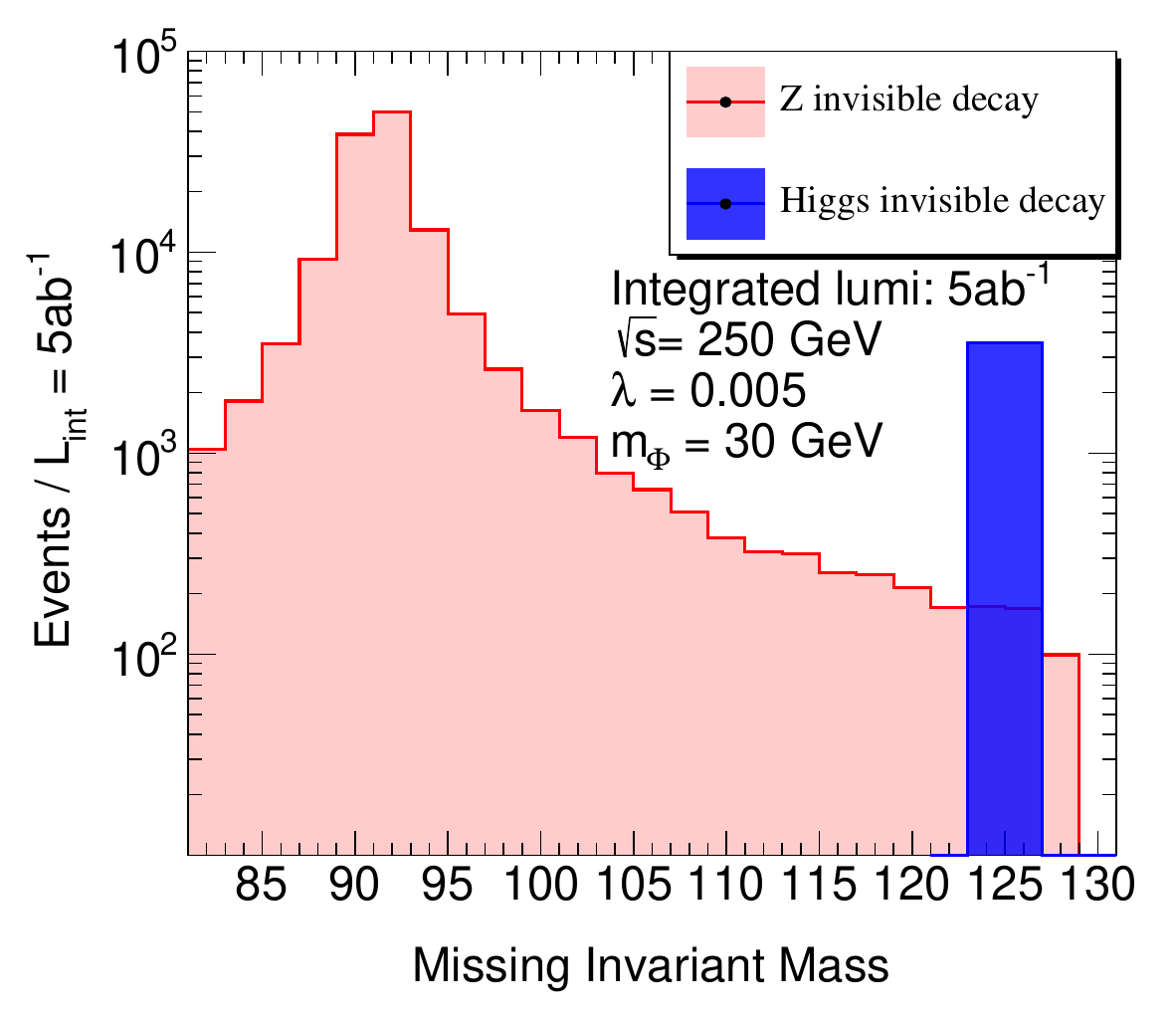}} 
	\subfigure[\label{fig:run5e330350}]
	{\includegraphics[width=0.45\textwidth]{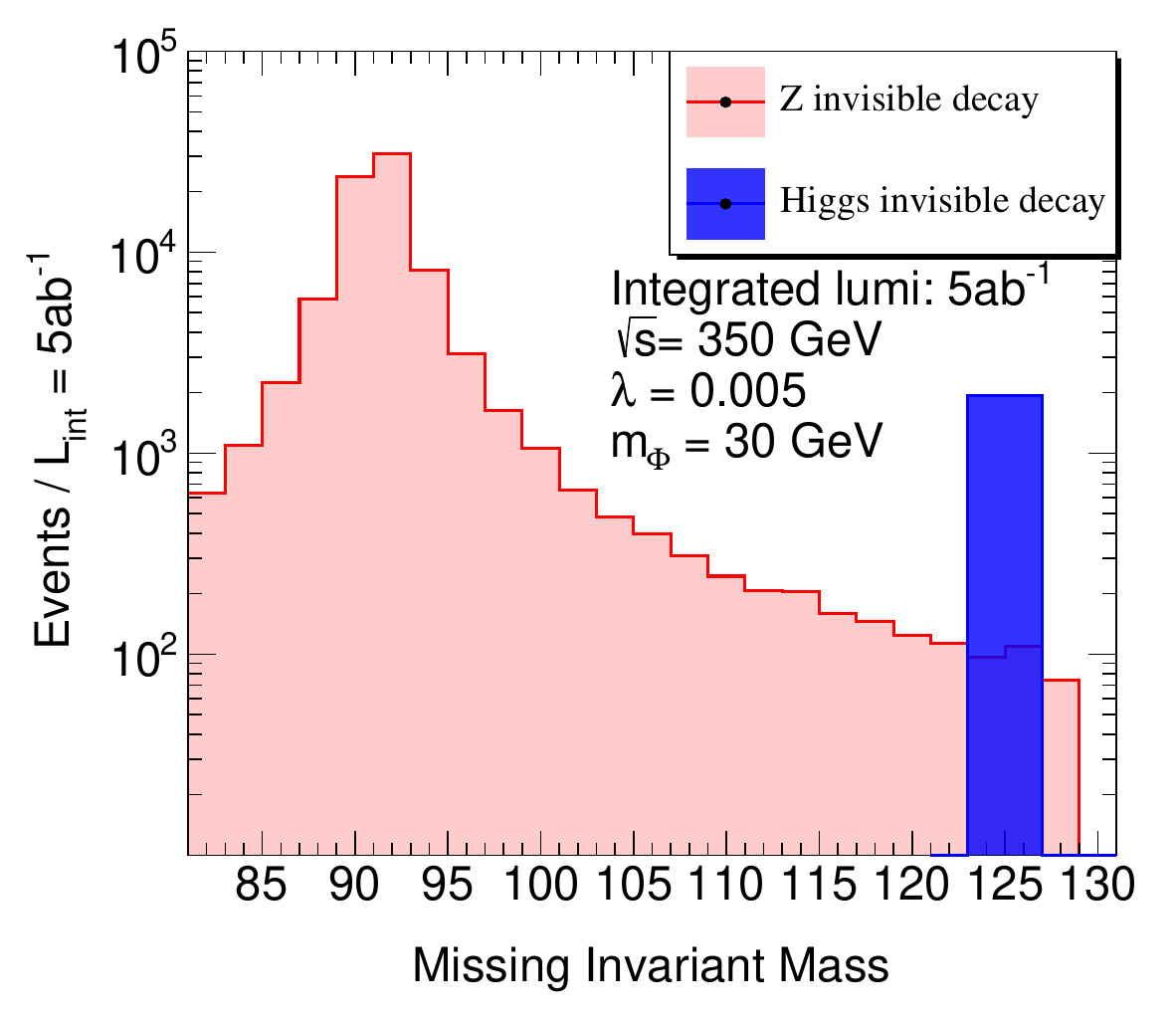}}  
	\caption{
	Missing invariant mass for different center-of-mass frame energy of the collider for BP1 are displayed: (a) $\sqrt{s} = 250$ GeV, around 7100 signal events and 350 background events between $125 \pm 2$ GeV (b) $\sqrt{s} = 350$ GeV, around 3850 signal events and 210 background events between $125 \pm 2$ GeV. Here we set the bin width to be 2 GeV, which is reasonable when the errors of the collision energy and momentum of charged leptons are considered.} 
	\label{fig:runINV}
\end{figure}

 As it is known that the total energy at the center-of-mass frame for $e^{+} e^{-}$ colliders can be determined with a remarkable precision (say 0.5 GeV due to the beam spreading) \cite{Behnke:2013lya}, therefore four momentum of the mother particles (say $H^0$ or $Z$ here) 
can be measured precisely in the central region of detectors
by accurate measurements of the four momentum of the final state lepton pair $p(l_1)$ and $p(l_2)$ from Z decay (say $\frac{\delta p }{p} = 10^{-5}$). 
By using these measurements, the four momentum of the invisible particle can be fully reconstructed via the so-called recoil mass method. 

The four momentum of the invisible mother particle denoted as $P^\mu_{inv} = q^\mu - P^\mu(Z)$ can be used to find its invariant mass, where $q^\mu=(\sqrt{s}, 0,0,0)$ and $P^\mu(Z) = p(\ell_1)^\mu + p(\ell_2)^\mu$. The invariant mass can be defined as $m_{\mathrm{inv}}^{\mathrm{miss}} = \sqrt{ P_{inv}^2 }$, which can be expressed as a function of collision energy and measured lepton four momenta 
\begin{equation}
	m_{\mathrm{inv}}^{\mathrm{miss}} = \sqrt{(\sqrt{s} - E_{l_1} - E_{l_2})^2-(\vec{p}(l_1)+\vec{p}(l_2))^2}\,.
\end{equation}
It should be stressed that the error of the observable $m_{\mathrm{inv}}^{\mathrm{miss}}$ only depends upon the error of $\sqrt{s}$ and the error of ($E_\ell$ and $p(\ell)$). For the signal events, the invariant mass $P^\mu_{inv}$ will yield a sharp peak at the Higgs mass region. In contrast, the events of the major background process $e^{+} e^{-} \rightarrow Z(\rightarrow l^{+} l^{-}) Z(\rightarrow \nu \nu)$ will lead to a peak at the Z mass region. 

Meanwhile, as it is well-known that the decay width of a Higgs boson is around 4 MeV, which is much narrower than that of a Z boson (2.49 GeV). Therefore, the signal events will enrich at a narrower Higgs boson mass window, i.e. $|m_{\mathrm{inv}}^{\mathrm{miss}} - m_H| \sim 2$ GeV, as demonstrated in Fig. \ref{fig:runINV} for BP1. Therefore, it clearly shows that it is possible to distinguish a signal/background event by setting a cut around the Higgs mass window $|m_{\mathrm{inv}}^{\mathrm{miss}}-125| = 2$ GeV. 

In Fig. \ref{fig:runINV} for BP1, even though the events via off-shell Z and $\gamma$ can contribute significantly even out of the $m_Z$ mass window, the signal events condense in the Higgs boson mass window, and a significance $\sigma=\frac{N_S}{\sqrt{N_B}} = 400$ for the BP1 can be established ($N_S$ and $N_B$ denote the number of signal and background events in the Higgs mass window ). For the collision energy $\sqrt{s}=350$ GeV, a significance $\sigma=265$ for the BP1 can be established. Compared with $\sqrt{s}=250$ GeV case, the loss of significance for $\sqrt{s}=350$ GeV can mainly be attributed to the fact that the cross section of the signal process is suppressed by collision energy since the signal process is s-channel dominant, while the cross section of background process is less suppressed since it is t/u-channel dominant.

\begin{figure}[!htbp]
	\centering
	\subfigure[\label{fig:run8e450250}]
	{\includegraphics[width=.45\textwidth]{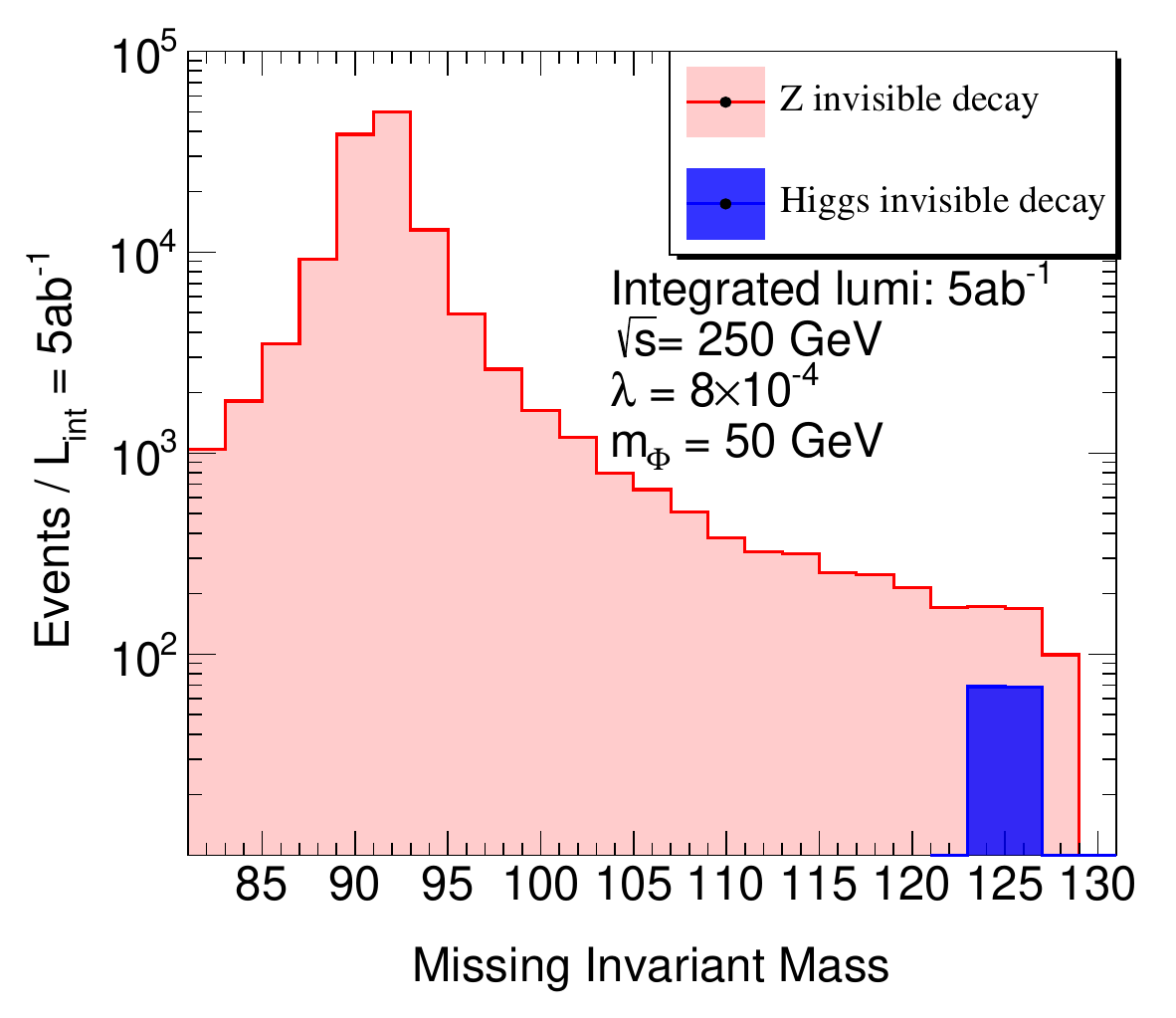}} 
	\subfigure[\label{fig:run8e450350}]
	{\includegraphics[width=0.45\textwidth]{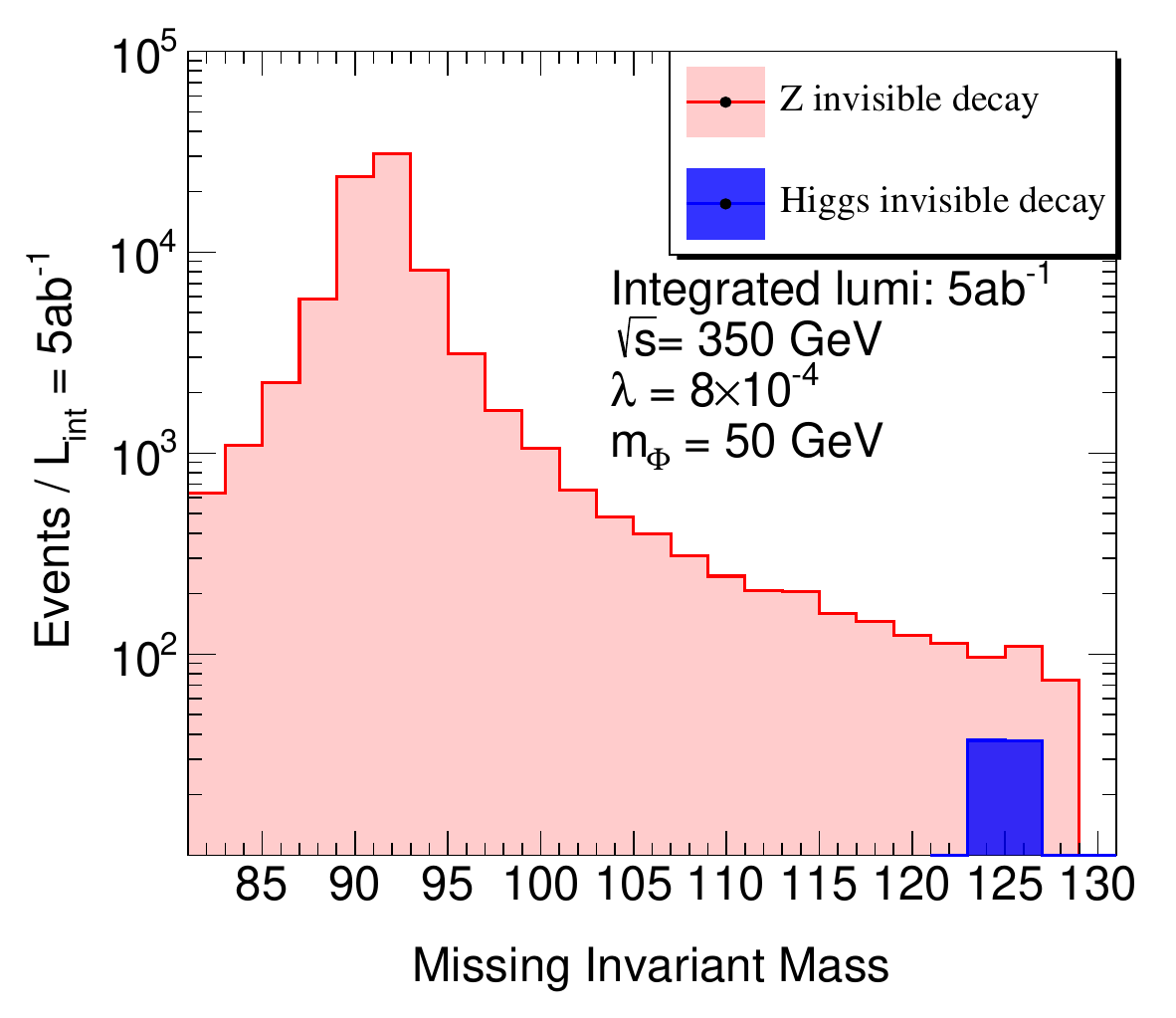}}  
	\caption{The recoiled mass for different center-of-mass frame energy of the collider for BP2 are displayed: (a) $\sqrt{s} = 250$ GeV, around 140 signal events and 350 background events between $125 \pm 2$ GeV. Significance of the signal is around 7.5 $\sigma$ (b) $\sqrt{s} = 350$ GeV, around 75 signal events and 210 background events between $125 \pm 2$ GeV. Significance of the signal is around 4.7 $\sigma$. Here we set the bin width to be 2 GeV, which is reasonable when the errors of the collision energy and momentum of charged leptons are considered.} 
	\label{fig:runINV50}
\end{figure}
 
In Fig.~\ref{fig:runINV50} for BP2, in the case with $\sqrt{s}=250$ GeV and an assumed 1/ab dataset, a significance $\sigma = 7.48$ can be achieved. While for the case with $\sqrt{s}=350$ GeV and the 1/ab dataset, a significance $\sigma = 4.70$ is feasible. For both BP1 and BP2, we can arrive at the conclusion that an $e^{+} e^{-}$ collider with $\sqrt{s}=250$ GeV has a better sensitivity for signal events. Note here that we can also choose the Higgs boson mass window to be $|m_{\mathrm{inv}}^{\mathrm{miss}}-125| = 1$ GeV. Comparing with the 2 GeV window, this scheme cut more background events and lead to a higher background rejection rate, so a higher significance can be obtained. While for the 2 GeV window, a higher signal efficiency will be achieved. 

We can extend our analysis for BP1 and BP2 to other points in the theoretical parameter space formed by $\lambda_{H \phi}$ and $m_\phi$ as demonstrated in Table. \ref{MincouplingEE}. We can obtain the lower bounds of the values of coupling $\lambda_{H \phi}$ 
which can be probed at $3 \sigma$ and $5 \sigma$ significance level for  integrated luminosity of both $\sqrt{s} = 250$ GeV and $\sqrt{s}=350$ GeV 
and for some different values of mass parameter $m_\phi$. 
\begin{table}[!htbp]
	\setlength{\tabcolsep}{2.8mm}{
		\begin{tabular}{|l|ll|ll|ll|}
%			\multicolumn{7}{c}{\textbf{Table 2:} }    
 \hline
			\multicolumn{7}{|c|}{$\sqrt{s} = 250$ GeV}  \\ \hline
			& \multicolumn{2}{|l|}{$m_{\phi}$ = 30 GeV} & \multicolumn{2}{|l|}{$m_{\phi}$  = 50 GeV} & \multicolumn{2}{|l|}{$m_{\phi}$ = 60 GeV} \\ \hline
			&  3 $\sigma$ & 5 $\sigma$  &  3 $\sigma$  &5 $\sigma$ & 3 $\sigma$  &5 $\sigma$     \\ \hline
			5 $ab^{-1}$  & $4.30 \times 10^{-4}$  & $5.60 \times 10^{-4}$ &   $5.20 \times 10^{-4}$ & $6.80 \times 10^{-4}$ & $7.60 \times 10^{-4}$ & $9.80 \times 10^{-4}$ \\ 
			1 $ab^{-1}$  & $6.60 \times 10^{-4}$ & $8.60 \times 10^{-4}$  &     $7.90 \times 10^{-4}$ & $1.03 \times 10^{-3}$ & $1.16 \times 10^{-3}$& $1.51 \times 10^{-3}$ \\ 
			500 $fb^{-1} $ & $7.90\times 10^{-4}$& $1.10 \times 10^{-3}$  & $9.50 \times 10^{-4}$ & $1.26 \times 10^{-3}$ & $1.40 \times 10^{-3}$& $1.84 \times 10^{-3}$ \\ \hline
			\multicolumn{7}{|c|}{$\sqrt{s} = 350$ GeV}  \\ \hline
			5 $ab^{-1}$  & $5.20 \times 10^{-4}$  & $6.70 \times 10^{-4}$ &   $6.30 \times 10^{-4}$ & $8.10 \times 10^{-4}$ & $9.20 \times 10^{-4}$ & $1.19 \times 10^{-3}$ \\ 
			1 $ab^{-1}$  & $7.90 \times 10^{-4}$ & $1.05 \times 10^{-3}$  &     $9.60 \times 10^{-4}$ & $1.26 \times 10^{-3}$ & $1.40 \times 10^{-3}$& $1.85 \times 10^{-3}$ \\ 
			500 $fb^{-1} $ & $9.60\times 10^{-4}$& $1.27 \times 10^{-3}$  & $1.15 \times 10^{-3}$ & $1.53 \times 10^{-3}$ & $1.71 \times 10^{-3}$& $2.23 \times 10^{-3}$ \\ \hline
		\end{tabular}
		\caption{The minimum values of coupling $\lambda_{H \phi}$ which can be probed at 3 $\sigma$ and 5 $\sigma$ confidence levels with different masses of $m_\phi$ with collision energy $\sqrt{s}=250$ GeV and  $\sqrt{s}=350$ GeV are shown.}
\label{MincouplingEE}
	}
\end{table}

When collision energies are fixed, the minimum integrated luminosity to obtain $3 \sigma$ and $5 \sigma$ significance for different values of the coupling $\lambda_{H\phi}$ and mass $m_\phi$ are shown in Table \ref{MinintEE}. 

From the results given in Table  \ref{MinintEE}, we can observe that the future $e^{+} e^{-}$ colliders can probe the signal process with the coupling $\lambda_{H \phi}$ down to $6 \times 10^{-4}$ or so. With the results of more points in theoretical parameter space, we can conclude that $e^+ e^-$ colliders with collision energy $\sqrt{s}=250$ can have a better sensitivity to our model.

\begin{table}[!htbp]
	\setlength{\tabcolsep}{0.8mm}{
		\begin{tabular}{|l|lll|lll|lll|}
\hline
			\multicolumn{10}{|c|}{$\sqrt{s} = 250$ GeV}  \\ \hline
			\multicolumn{1}{|c|}{} & \multicolumn{3}{|l|}{$m_{\phi} = 30$ GeV} & \multicolumn{3}{|l|}{$m_{\phi} = 50$ GeV} & \multicolumn{3}{|l|}{$m_{\phi} = 60$ GeV} \\ \hline
			$\lambda_{H \phi}$  & $6 \times 10^{-4} $  & $8\times 10^{-4}$     & $1 \times 10^{-3} $   & $6 \times 10^{-4}$ & $8 \times 10^{-4}$  & $1 \times 10^{-3}$   & $8 \times 10^{-4} $  & $1\times 10^{-3}$     & $1.2 \times 10^{-3} $  \\
			3 $\sigma$  &  1.41 $ab^{-1}$  &   480 $fb^{-1}$  &  230 $fb^{-1}$   &   2.90  $ab^{-1}$  &  940 $fb^{-1}$  & 430 $fb^{-1}$   &   4.10  $ab^{-1}$  &  1.70 $ab^{-1}$  & 890 $fb^{-1}$ \\
			5 $\sigma$  &   3.80  $ab^{-1}$ &  1.27 $ab^{-1}$ & 560 $fb^{-1}$ &  7.80 $ab^{-1}$ & 2.60 $ab^{-1}$ & 1.13 $ab^{-1}$  & 11.20 $ab^{-1}$ & 4.70 $ab^{-1}$ & 2.38 $ab^{-1}$ \\ \hline  
			\multicolumn{10}{|c|}{$\sqrt{s} = 350$ GeV}  \\ \hline
			$\lambda_{H \phi}$  & $6 \times 10^{-4} $  & $8\times 10^{-4}$     & $1 \times 10^{-3} $   & $6 \times 10^{-4}$ & $8 \times 10^{-4}$  & $1 \times 10^{-3}$   & $8 \times 10^{-4} $  & $1\times 10^{-3}$     & $1.2 \times 10^{-3} $  \\
			3 $\sigma$  &  2.9 $ab^{-1}$  &  990 $fb^{-1}$  &  420 $fb^{-1}$   &   5.80  $ab^{-1}$  &  1.97 $ab^{-1}$  & 860 $fb^{-1}$   &   8.40  $ab^{-1}$  &  3.60 $ab^{-1}$  & 1.80 $ab^{-1}$ \\
			5 $\sigma$  &   7.85  $ab^{-1}$ &  2.70 $ab^{-1}$ & 1.16 $ab^{-1}$ &  16.10 $ab^{-1}$ & 5.30 $ab^{-1}$ & 2.31 $ab^{-1}$  & 23.00 $ab^{-1}$ & 9.80 $ab^{-1}$ & 4.80 $ab^{-1}$ \\ \hline  
		\end{tabular}
\caption{The minimum integrated luminosity for 3 $\sigma$ and 5 $\sigma$ confidence level for different masses $m_\phi$ and values of coupling $\lambda_{H \phi}$ with two collision energy $\sqrt{s}=250$ GeV and $\sqrt{s}=350$ GeV  are provided.}
	\label{MinintEE}}
\end{table}

\section{Dark sector particle $\phi$ search at LHC}
\label{ppcollider}
For the sake of comparison, we study the feasibility to detect dark particle $\phi$ at proton-proton colliders. We focus on both mono-jet and $2 \ell + E_{\mathrm{T}}^{\mathrm{miss}}$ final states at pp colliders. 

%\textcolor{red}{hadronization, jet algorithm}

It is well-known that the mono-jet process serves as a unique probe to search for the DM production processes which lead to a large $E_{\mathrm{T}}^{\mathrm{miss}}$ and one energetic jet in the final state. In our model, the signal process is $pp \to H + j$ with Higgs boson decaying into a $\phi$ pair. The dominant background process is $pp \to Z + j$ with the Z boson decaying into a neutrino pair. The detailed signal/background processes and decays are given below 

\begin{equation}
\begin{aligned}
&p p  \rightarrow H+\text{jet} \rightarrow E_{\mathrm{T}}^{\text {miss }}(\phi \phi)+\text{jet} \quad (\textbf{signal}) \\
&p p \rightarrow Z +\text{jet} \rightarrow E_{\mathrm{T}}^{\text {miss }}(\nu \bar{\nu})+\text{jet} \ \quad (\textbf{background})
\end{aligned}
\end{equation}

We generate the relevant the mono-jet events at parton level by using \textbf{MadGraph5\_aMC@NLO}\cite{Alwall:2011uj,Alwall:2014hca} generator for signal and background processes. 
%%%%%%
For simplicity, only one energetic jet in the final state is considered. The PDF set used for the generation is NNPDF23LO, and the cross section for the process with the hard jet in the final state is obtained by using a dataset at parton level (no hadronization or no detector effects has been considered) .
We set the following selection cuts to the leading jet: $p_{T}(j) > 150$ GeV and $|\eta(j)|<2.4$.
%%%%%%

\begin{figure}[!t]
	\centering
	\subfigure[\label{fig:Monojet5e330}]
	{\includegraphics[width=.486\textwidth]{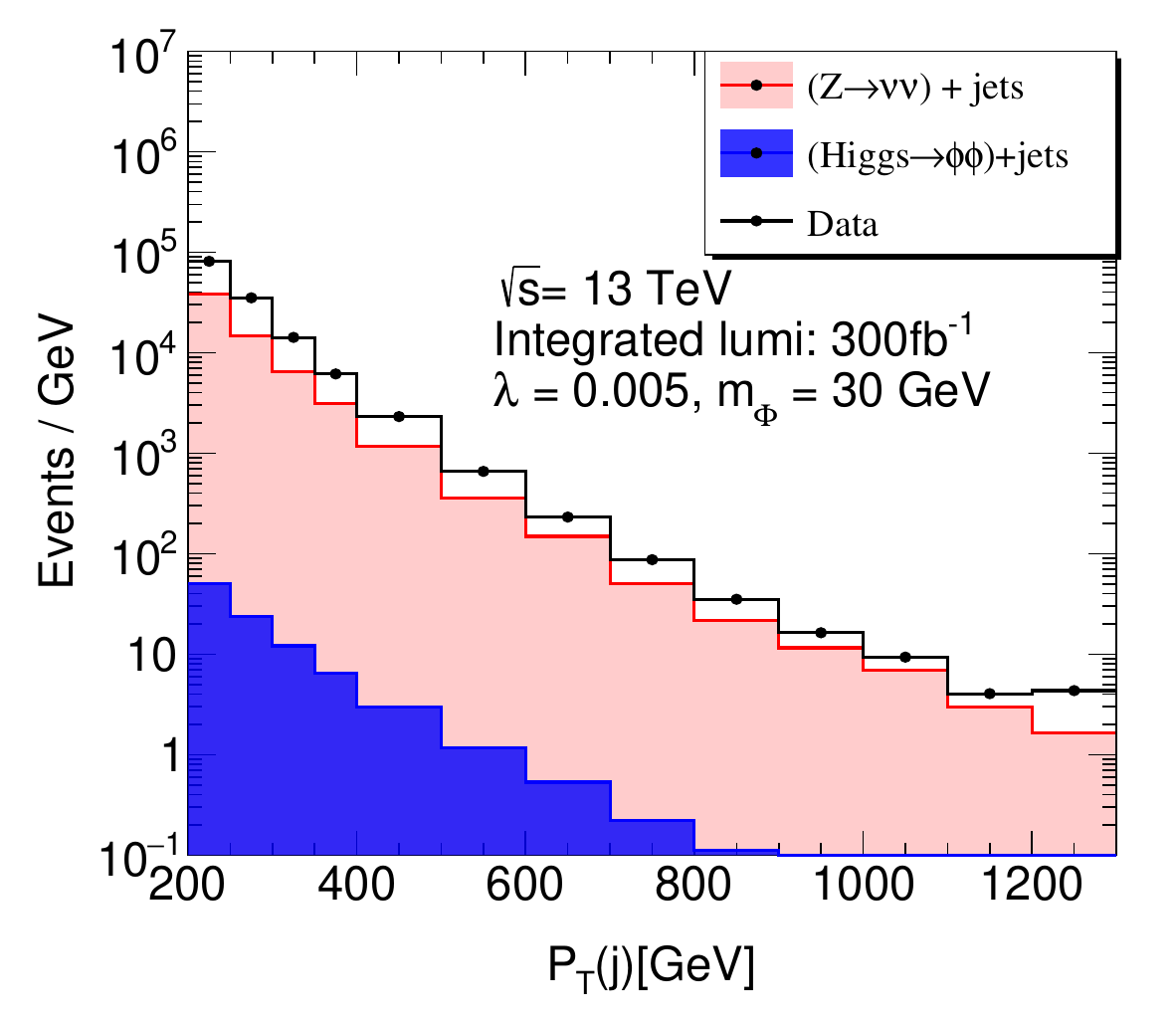}} 
	\subfigure[\label{fig:Monojet5e350}]
	{\includegraphics[width=0.486\textwidth]{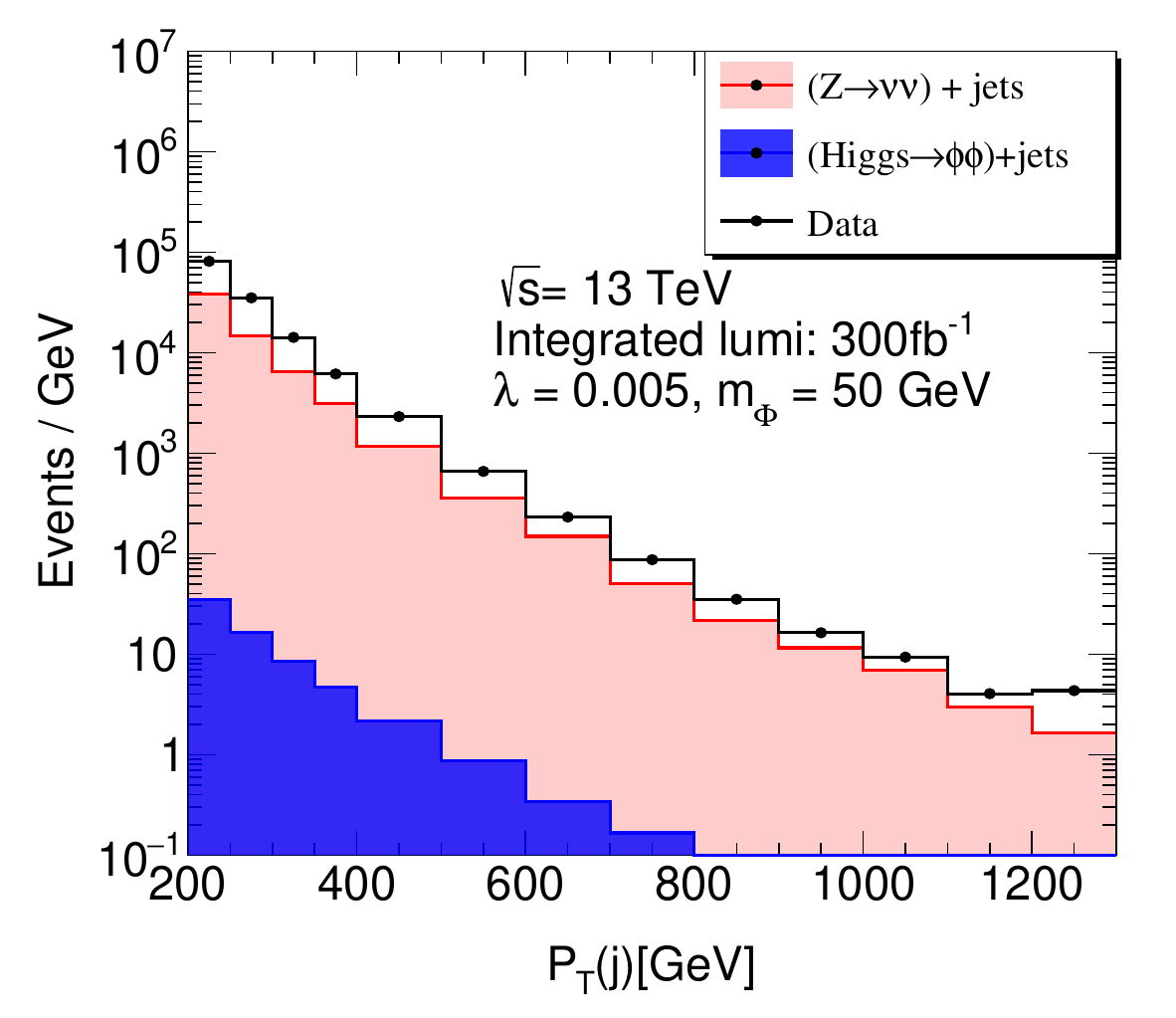}}  
	\caption{(a) $p_T(j)$ distribution of signal events and background events in the monojet processes. The black points represent the experimental data. (a)$\lambda_{H \phi} = 5 \times 10^{-3}$, $m_{\phi} = 30$ GeV. (b)$\lambda_{H \phi} = 5 \times 10^{-3}$, $m_{\phi} = 50$ GeV.} 
	\label{fig:Monojet}
\end{figure}

In Fig.~\ref{fig:Monojet}, the distributions of the transverse momentum of the leading jet $p_{T}(j)$ of signal/background events are shown. For the signal events, we have chosen the theoretical parameters $m_{\phi} = 30 $ GeV in the left panel and $m_{\phi} = 50 $ GeV in the right panel with $\lambda_{H \phi} = 5 \times 10^{-3}$. The blue(red) bins represent the distribution of signal(background) events. 

To examine the reliability of our MC simulation, the experimental data are denoted by the black points, which is normalized to the experimental results for mono-jets search at LHC with integrated luminosity $139 fb^{-1}$\cite{ATLAS:2021enr,ATLAS:2021kxv} and simply rescaled to $300 fb^{-1}$. It is noticed that there exists a small gap between the experimental data and the results of MC simulation which can be attributed to the W+jets processes and the NLO contribution, as has been taken into account by \cite{ATLAS:2021enr}. 

We can find in both Fig.~\ref{fig:Monojet5e330} and Fig.~\ref{fig:Monojet5e350} that the background events are approximately 3 orders of magnitude larger than signal events for each a bin, even when the NLO correction for the $gg \to H^0 j$ is taken into account. This indicates that it is really challenging for the LHC to find the signal process from the mono-jets events with a total integrated luminosity $300 fb^{-1}$ after run3. 

\begin{table}[htbp!]
	\centering
	\label{PPcut}
	\setlength{\tabcolsep}{2.8mm}{
		\begin{tabular}{lllllllll}
			\hline
			\multicolumn{9}{c}{Selection cuts for mono-jet final states}    \\ \hline
			\multicolumn{3}{l}{Leading jet} & \multicolumn{6}{l}{$p_{T}(j) > 150$ GeV and $|\eta|<2.4$} \\ \hline \hline
			\multicolumn{9}{c}{Selection cuts for $ ZH \rightarrow \ell \ell+E_{\mathrm{T}}^{\text {miss}}$ final states}  \\ \hline 
			\multicolumn{3}{l}{Two leptons} & \multicolumn{6}{l}{leading(subleading) $p_{T}(\ell)$ > 30(20) GeV, $|\eta(\ell)|<2.5$} \\
			\multicolumn{3}{l}{$m_{\ell \ell}$} & \multicolumn{6}{l}{$76$ GeV $<m_{\ell \ell}<106$ GeV} \\
			\multicolumn{3}{l}{$E_{\mathrm{T}}^{\text {miss }}$} & \multicolumn{6}{l}{$E_{\mathrm{T}}^{\text {miss }} >  90 $GeV} \\
			\multicolumn{3}{l}{$\Delta R_{\ell}$} & \multicolumn{6}{l}{ $\Delta R_{\ell} \geq 0.4 $}\\ \hline
		\end{tabular}	
		\caption{Selection cuts for mono-jet and $2\ell + E_{\mathrm{T}}^{\text {miss }}$ final states at pp colliders are presented. }
}
\end{table}

The process for production of Higgs boson in association with a Z boson with the Higgs boson invisibly decay is also a good channel to search for new physics at pp colliders. 
The signal process and major background process is similar with the $e^{+} e^{-}$ collider case with proton instead of electron in the initial state.

\begin{equation}
\begin{aligned}
&p p \rightarrow ZH \rightarrow l^{+} l^{-} + E_{\mathrm{T}}^{\mathrm{miss}}(\phi \phi) \quad (\textbf{signal})\\
&p p \rightarrow ZZ \rightarrow l^{+} l^{-} + E_{\mathrm{T}}^{\mathrm{miss}}(\nu \bar{\nu}) \ \quad (\textbf{background})
\end{aligned}
\end{equation}
Since we can not reconstruct the full momentum of the missing particles in the final state at pp colliders, instead we can only use the transverse missing momentum  $E_{\mathrm{T}}^{\mathrm{miss}}$ as an observable.

\begin{figure}[!t]
	\centering
	\subfigure[\label{fig:ZHLL5e330}]
	{\includegraphics[width=.486\textwidth]{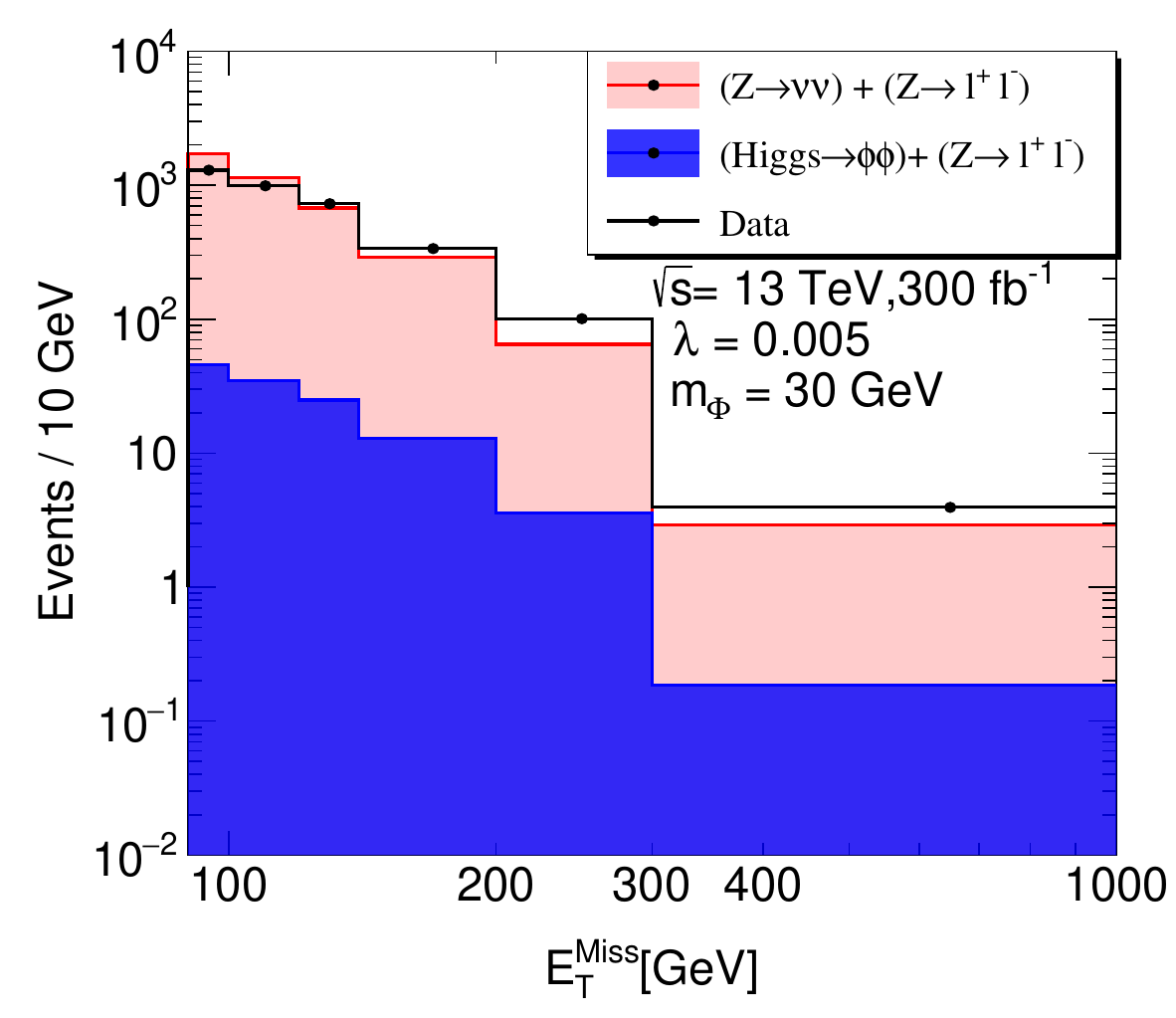}} 
	\subfigure[\label{fig:ZHLL5e350}]
	{\includegraphics[width=0.486\textwidth]{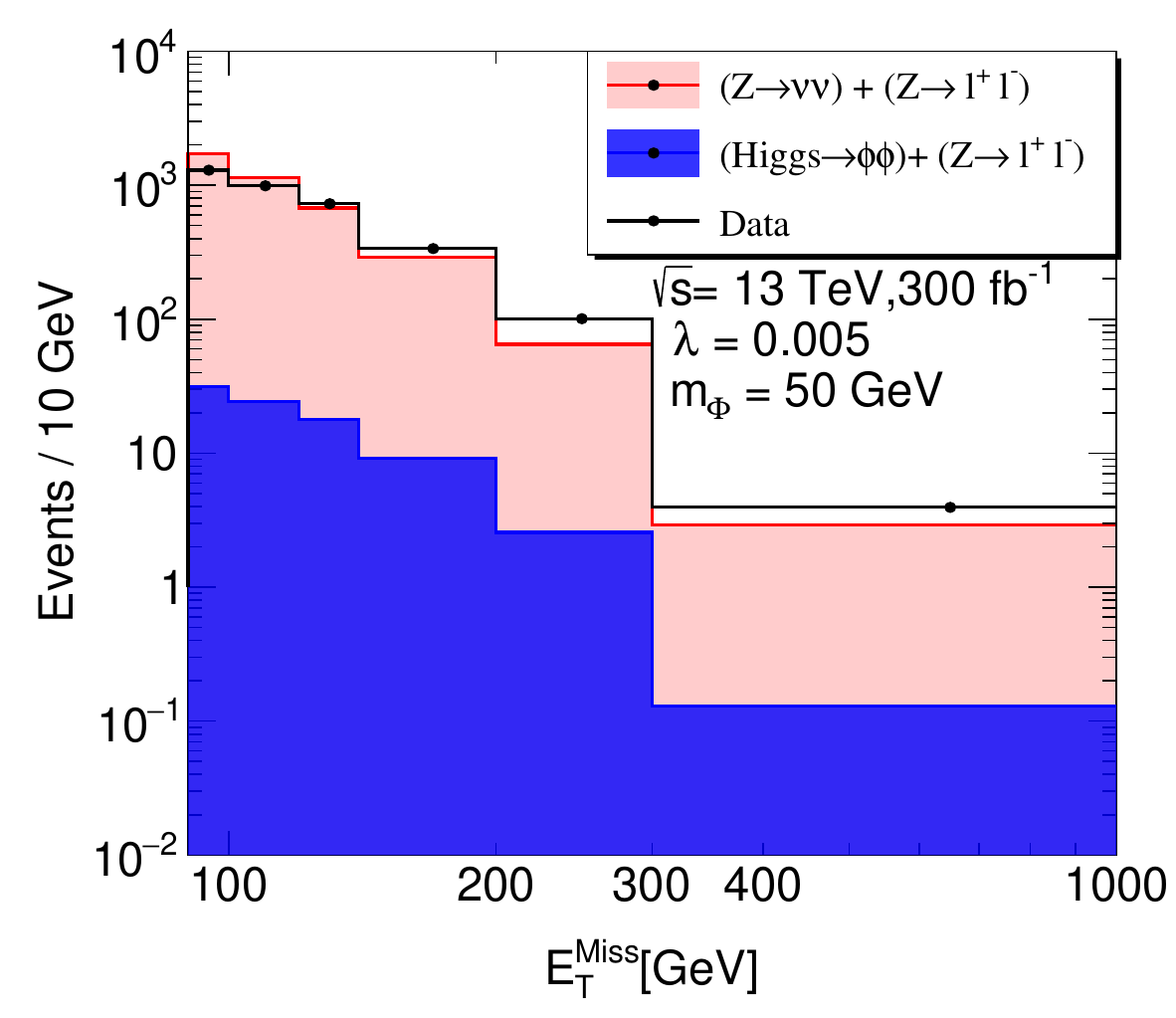}}  
	\caption{Distribution of $E_{\mathrm{T}}^{\mathrm{miss}}$ for signal process $(H \rightarrow \phi \phi)$ + $(Z\rightarrow l^{+} l^{-})$ and background process $(Z \rightarrow \nu  \bar{\nu})$ + $(Z\rightarrow l^{+} l^{-})$. The black points represent the experimental data. (a)$\lambda_{H \phi} = 5 \times 10^{-3}$, $m_{\phi} = 30$ GeV. (b)$\lambda_{H \phi} = 5 \times 10^{-3}$, $m_{\phi} = 50$ GeV.}
	\label{fig:PPZHLL}
\end{figure}

In Fig.~\ref{fig:PPZHLL}, we show the distribution of $E_{\mathrm{T}}^{\mathrm{miss}}$ for signal/background in blue/red at LO and compare the experimental data with the Monte Carlo results. Here the ATLAS results for invisibly decaying Higgs boson in association with a Z boson data at 36.1 $fb^{-1}$\cite{ATLAS:2017nyv} are quoted and scaled to 300 $fb^{-1}$. 

In our analysis, we demand that the leptons $p_{T}(\ell) > 20$ GeV and the leading lepton $p_T(\ell) > 30$ GeV. Also $E_{\mathrm{T}}^{\mathrm{miss}} > 90 \mathrm{GeV}$ is demanded and the invariant mass of two charged leptons is required to be in the range $76$ GeV $< m_{\ell \ell} < 106 ~\mathrm{GeV}$ 
 to reject background processes with two leptons that do not originate from the decay of a Z boson. All the cuts we use are shown in Table \ref{PPcut}. 

Note here that the result has a deviation of about 10\% from the experimental data where the NLO corrections have been taken into account. We can find in the figure that the signal is 2 orders of magnitude smaller than the background. So the signal events for dark sector particle $\phi$ is also very hard to be found from ZH process at pp colliders.

Comparing these results with those at $e^{+} e^{-}$ colliders given in Sec.~\ref{eecollider}, we can conclude that $e^+e^-$ colliders have more advantages in searching for such dark sector particle $\phi$ from Higgs invisible decay.

\section{Conclusions \label{sec:conc}}

In summary, we continued the research on the scenario that the DM  can be produced by the decay of other heavier particles in the dark sector. For the specific model which include one singlet scalar $\phi$ and one singlet fermion $\chi$ in the dark sector, we have studied the possibility for a lighter dark sector particle $\phi$ with $m_{\phi} < 100$ GeV to obtain the right relic density. We find this scenario with lower $m_{\phi}$
 is also appropriate in  a wide range of parameter space.

We have studied the possibility to find the signal of  the dark sector particle $\phi$ on colliders. 
We mainly focus on the region that $m_{\phi} < m_{H}/2$ and have  studied the production rate and possible signals on the colliders. 
For future Higgs factory running at 250-350 GeV, the dark particle $\phi$ is mainly produced from ZH process and will result in the dilepton final state with large $E_{\mathrm{T}}^{\mathrm{miss}}$. This kind of signature can also be generated by SM process from diboson(ZZ) decay, which is the domant background events. From reconstructing the full missing three momentum and energy we can obtain the invariant mass of the missing part which give us a powerful tool to distinguish the signal and background events. Using the missing invariant mass as an observable, we can probe this kind of dark sector signal with the coupling of $\phi$ and Higgs down to $5\times10^{-4}$ at future Higgs factories. We have studied the signal and background processes for some benchmark points in our model and have found that the signal has strong significance for a wide range of the parameter space.
We have also studied the sensitivity for searching this dark sector particle signal at Higgs factories with different center-of-mass frame energy. 
We find that a center-of-mass frame energy 250 GeV is slightly better to find the dark sector signal.

For comparison, we have also studied the possibility of looking for the same dark sector signals on the pp colliders. Higgs production and then decay to $\phi$ pair associated with an energetic jet in our model will lead to the Mono-jet signals, for which $Z \rightarrow \nu \bar{\nu}+$ jets process is the major SM background. We have studied $p_T(j)$ distribution for signal and background and have found that the signal is 3 orders of magnitude smaller the background. 
We have further studied the dark sector particle produced by ZH process at pp colliders, where $(ZZ \rightarrow  l^{+} l^{-})+E_{\mathrm{T}}^{\mathrm{miss}}$ is the major background in the SM. Since we can not reconstruct the full missing momentum, we could only choose transverse missing energy as an observable. We find that for some benchmark points of the model, the signal is 2 orders of magnitude smaller than the background in the $E_{\mathrm{T}}^{\mathrm{miss}}$ plot. We find that for the missing energy signals considered in this article
the pp collider is not a good place to detect the signal of the dark sector $\phi$ particle.

Note that the dark sector particle $\phi$ may lead to signals of displaced vertex in the detector, which depends upon the dark sector and right-handed neutrinos coupling $y_{DS}$. We will leave this topic in the future work.

\bigskip
\section*{Acknowledgements}
\label{sec:acknowledgements}
W. Liao is supported by National Natural Science Foundation of China under the grant No. 11875130.
Q.S.Yan is supported by the National Natural Science Foundation of China under the grant No. 11475180 and  No. 11875260.

%\bibliography{darkmatter}

%\bibliographystyle{utphys}

%
\providecommand{\href}[2]{#2}\begingroup\raggedright\endgroup

\end{document}